\documentclass[conference]{IEEEtran}
\IEEEoverridecommandlockouts
\usepackage{cite}
\usepackage{amsmath,amssymb,amsfonts}
\usepackage{algorithmic}
\usepackage{graphicx}
\usepackage{textcomp}
\usepackage{xcolor}
\usepackage{array}

\usepackage{soul}
\setstcolor{red}

\newcounter{tempcounter_1}
\newcounter{tempcounter_2}
\makeatletter
\newcommand{\linebreakand}{%
  \end{@IEEEauthorhalign}
  \hfill\mbox{}\par
  \mbox{}\hfill\begin{@IEEEauthorhalign}
}
\makeatother

\def\BibTeX{{\rm B\kern-.05em{\sc i\kern-.025em b}\kern-.08em
    T\kern-.1667em\lower.7ex\hbox{E}\kern-.125emX}}
    
\begin{document}

\title{Transferable Selective Virtual Sensing Active Noise Control Technique Based on Metric Learning
\thanks{The code will be available at https://github.com/Wang-Boxiang/Transferable-Selective-Virtual-Sensing-Active-Noise-Control}
}





\author{
\IEEEauthorblockN{1\textsuperscript{st} Boxiang Wang}
\IEEEauthorblockA{
\textit{Nanyang Technological University}\\
Singapore \\
boxiang001@e.ntu.edu.sg}
\and
\IEEEauthorblockN{2\textsuperscript{nd} Dongyuan Shi}
\IEEEauthorblockA{
\textit{Nanyang Technological University}\\
Singapore \\
dongyuan.shi@ntu.edu.sg}
\and
\IEEEauthorblockN{3\textsuperscript{rd} Zhengding Luo}
\IEEEauthorblockA{
\textit{Nanyang Technological University}\\
Singapore \\
luoz0021@e.ntu.edu.sg}
\linebreakand
\IEEEauthorblockN{4\textsuperscript{th} Xiaoyi Shen}
\IEEEauthorblockA{
\textit{Nanyang Technological University}\\
Singapore \\
xiaoyi.shen@ntu.edu.sg}
\and
\IEEEauthorblockN{5\textsuperscript{th} Junwei Ji}
\IEEEauthorblockA{
\textit{Nanyang Technological University}\\
Singapore \\
junwei002@e.ntu.edu.sg}
\and
\IEEEauthorblockN{6\textsuperscript{th} Woon-Seng Gan}
\IEEEauthorblockA{
\textit{Nanyang Technological University}\\
Singapore \\
ewsgan@ntu.edu.sg}
}
\maketitle

\begin{abstract}
Virtual sensing (VS) technology enables active noise control (ANC) systems to attenuate noise at virtual locations distant from the physical error microphones. Appropriate auxiliary filters (AF) can significantly enhance the effectiveness of VS approaches. The selection of appropriate AF for various types of noise can be automatically achieved using convolutional neural networks (CNNs). However, training the CNN model for different ANC systems is often labour-intensive and time-consuming. To tackle this problem, we propose a novel method, Transferable Selective VS, by integrating metric-learning technology into CNN-based VS approaches. The Transferable Selective VS method allows a pre-trained CNN to be applied directly to new ANC systems without requiring retraining, and it can handle unseen noise types. Numerical simulations demonstrate the effectiveness of the proposed method in attenuating sudden-varying broadband noises and real-world noises.
\end{abstract}


\vspace*{-0.4cm}
\section{Introduction}
\vspace*{-0.2cm}
Active noise control (ANC) is an advanced approach to attenuating unwanted noise based on the principle of sound-destructive superposition~\cite{nelson1991active,elliott1993active,hansen1999understanding}. In an ANC system, the control filter derives the secondary source to generate the anti-noise with the same amplitude and opposite phase to the disturbance, so the noise pressure at the location of the error microphone can be minimized \cite{NNZD}. To react to the noise and environment change, the filtered reference least mean square (FxLMS) algorithm is often used to adaptively update the control filter coefficients~\cite{qiu2001study,morgan2013history,yang2020stochastic,7324088,ji2023practical}. Compared to the passive approaches, the ANC technique is particularly efficient in tackling low-frequency noises~\cite{kuo1999active}. Therefore, ANC is widely applied in various areas, such as headphones~\cite{shen2022adaptive,shen2022multi}, headrests~\cite{jung2019local,shen2023implementations} and windows~\cite{lam2018active,lam2020active,shi2023computation}.

However, as the noise frequency increases, the zone of quietness (ZoQ) created by the ANC system significantly diminishes, becoming confined to the area surrounding the error microphone. As a result, noise reduction performance at the target area can be compromised when the error microphone cannot be placed due to physical constraints. For instance, in an ANC headrest, it is impractical to place the error microphone near the user’s eardrum, which limits the formation of a ZoQ within the ear.
To address this issue, several virtual sensing (VS) techniques have been proposed~\cite{pawelczyk2004adaptive,moreau2008review,miyazaki2015head,jung2017combining,shi2020active,misol2020active,shi2020vs,sun2022spatial,wang2024computation,shi2025virtual}. VS techniques enable the creation of a ZoQ at desired virtual locations distant from the physical error microphone. Among these techniques, the auxiliary-filter-based VS (AF-VS) approach embeds the optimal noise control filter information into the auxiliary filter (AF) during the tuning stage and then uses this information to estimate the virtual error signal during the control stage~\cite{pawelczyk2004adaptive,miyazaki2015head,shi2020vs,wang2024computation}. Nevertheless, conventional AF-VS techniques perform noticeably worse when noise characteristics differ between the tuning and control stages. To address this issue, Shi et al. introduced the selective VS technique, inspired by selective fixed-filter active noise control (SFANC) systems~\cite{shi2020feedforward,shi2022selective,luo2022hybrid,luo2024real}, to accommodate variations in noise characteristics between these stages using a signal processing approach~\cite{shi2019selective}. Building on this concept, Xie et al. developed the cognitive VS technique, which utilizes a convolutional neural network (CNN) to account for changes in noise characteristics and primary paths~\cite{xie2024cognitive}.

Despite these advancements, various VS systems often encounter differing acoustic paths and noise types, necessitating distinct pre-trained AFs and CNN models. Deploying the CNN model across different systems requires collecting different noise datasets and retraining, which is both challenging and time-intensive. Moreover, the CNN model is prone to overfitting if the dataset is not sufficiently large. These practical challenges limit the widespread adoption of CNN-based VS techniques. A study on SFANC systems, excluding the VS technique, has shown that metric learning can enhance the generality of pre-trained CNN models~\cite{shi2023transferable}. Building on insights from this study, this paper proposes a transferable selective VS technique designed to address the challenges posed by varying noise characteristics across different VS systems. To classify noise in different acoustic environments, we retain the convolutional layers of the pre-trained CNN model and utilize a similarity-matching method based on metric learning. The proposed method can handle noise classes not present in the training set, allowing the pre-trained convolutional layers to be applied in new ANC platforms without retraining. The applied CNN model features a simple structure while achieving satisfactory noise classification accuracy with few parameters and computational requirements. Numerical simulations of the proposed transferable selective VS technique applied to cancel sudden-varying broadband and real-world noise validate its effectiveness in practical applications.

\vspace*{-0.2cm}
\section{CNN-based Selective Virtual Sensing Technique}
\vspace*{-0.25cm}
The conventional AF-VS technique consists of two stages: the tuning stage and the control stage. In the tuning stage, the error microphone is temporarily placed at the desired noise reduction area, and the $z$ domain representation of the virtual error signal ${E_v}(z)$ can be expressed as
\begin{equation}
\setlength{\abovedisplayskip}{4pt}
\setlength{\belowdisplayskip}{4pt}
{E_v}(z) = {P_v}(z)X(z) - {W_\mathrm{opt}}(z){S_v}(z)X(z),
\label{Eq:1}
\end{equation}
where $X(z)$ and ${W_\mathrm{opt}}(z)$ represent the reference signal and control filter in the tuning stage; ${{P_v}(z)}$ and ${{S_v}(z)}$ represent the virtual primary path and virtual secondary path, respectively. 

To minimize the virtual error signal, the optimal control filter can be derived as 
\begin{equation}
\setlength{\abovedisplayskip}{6pt}
\setlength{\belowdisplayskip}{6pt}
W_{opt}^o(z) = \frac{{{P_v}(z)}}{{{S_v}(z)}}.
\label{Eq:2}
\end{equation}

At this moment, the error signal at the physical error microphone ${E_p}(z)$ is given by
\begin{equation}
\setlength{\abovedisplayskip}{5pt}
\setlength{\belowdisplayskip}{5pt}
{E_p}(z) = {P_p}(z)X(z) - W_\mathrm{opt}^\mathrm{o}(z){S_p}(z)X(z),
\label{Eq:3}
\end{equation}
where ${{P_p}(z)}$ represents the physical primary path, and ${{S_p}(z)}$ represents the physical secondary path.

The optimal AF ${H^\mathrm{o}}(z)$ can be derived as 
\begin{equation}
\setlength{\abovedisplayskip}{5pt}
\setlength{\belowdisplayskip}{5pt}
{H^\mathrm{o}}(z) = {P_p}(z) - \frac{{{S_p}(z){P_v}(z)}}{{{S_v}(z)}} = {P_p}(z) - {S_p}(z)W_\mathrm{opt}^\mathrm{o}(z).
\label{Eq:4}
\end{equation}

In the control stage, the virtual error microphone is removed, leaving only the physical error microphone.  The trained AF then aids in training the new control filter $W(z)$ by minimizing the difference between the physical error signal and the AF output, represented by ${E_h}(z)$ as 
\begin{equation}
\setlength{\abovedisplayskip}{5pt}
\setlength{\belowdisplayskip}{5pt}
    \begin{aligned}
        {E_h}(z) &= {E_p}^\prime (z) - {H^\mathrm{o}}(z)X'(z) \\
        &= {S_p}(z)[W_\mathrm{opt}^\mathrm{o}(z) - W(z)]X'(z),
    \end{aligned}
    \label{Eq:5}
\end{equation}
where $X'(z)$ and ${E_p}^\prime (z)$ denote the reference signal and physical error signal in the control stage, respectively. The FxLMS algorithm converges as ${E_h}(z)$ approaches zero, resulting in a control filter nearly identical to the optimal one. 

\begin{figure}[htb]
\centering
\centerline{\includegraphics[width=6.5cm]{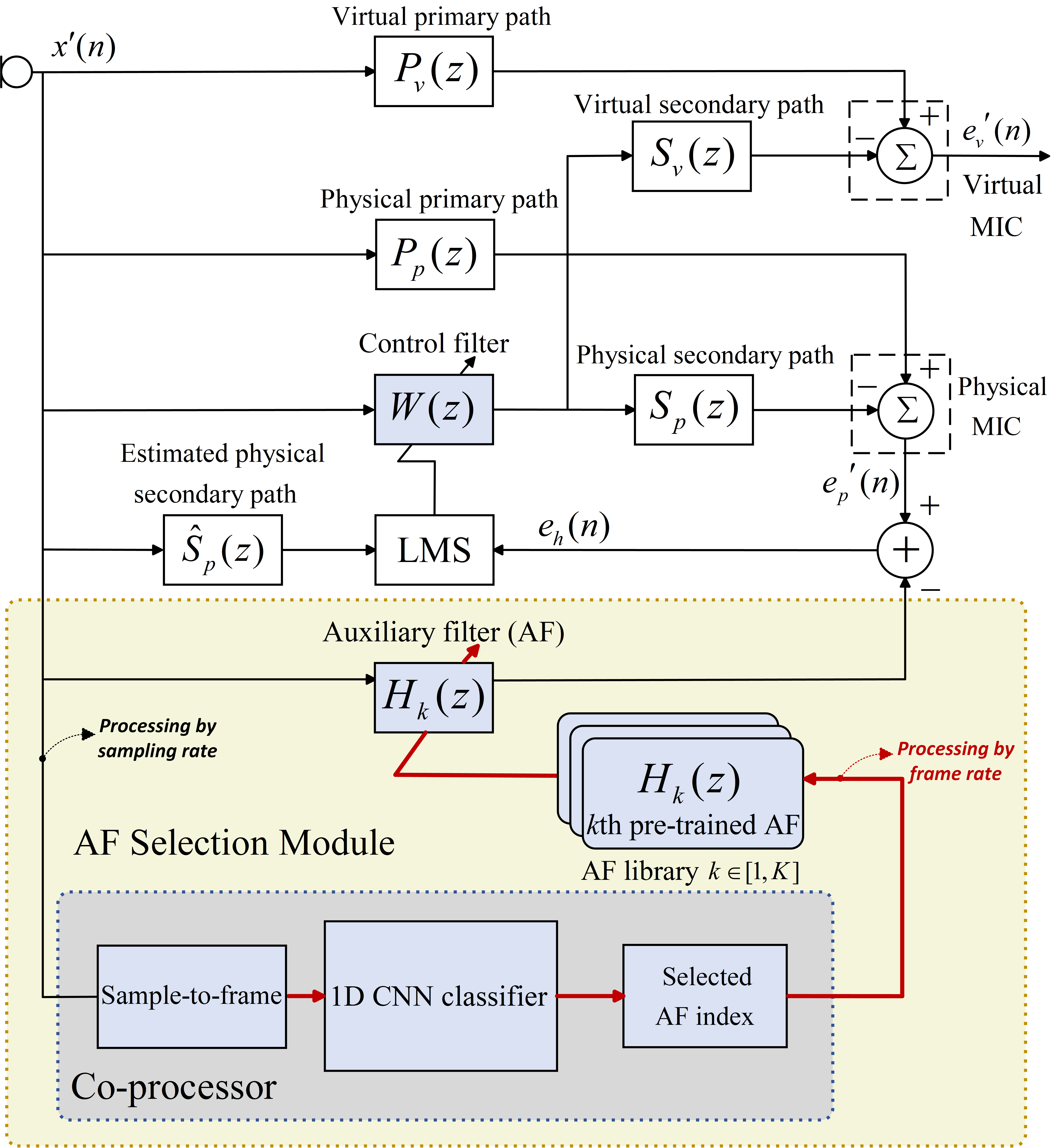}}\vspace*{-0.1cm}
\caption{Block diagram of the CNN-based selective virtual sensing technique.}
\label{fig:1}\vspace*{-0.5cm}
\end{figure}

It is noteworthy that the primary noise in the tuning stage differs from that in the control stage. Equation \eqref{Eq:5} indicates that the frequency band of the reference signal affects the training of the new control filter. Ideally, when the reference signal's frequency band in the tuning stage is wider than that in the control stage, the training of the new control filter remains unaffected. However, in real-world scenarios, additional noise captured by the reference microphone and acoustic paths can degrade the effectiveness of the AF-VS technique, particularly when the AF is trained for full-band. Therefore, the AF must be retrained when the primary noise's frequency band changes.

To overcome the limitations of the conventional AF-VS technique, the CNN-based selective VS technique is proposed, which dynamically switches the AF based on the characteristics of the primary noise encountered in the control stage. During the tuning stage, $K$ different AFs are independently trained using primary noises with various frequency bands and stored in a library. As illustrated in Fig.~\ref{fig:1}, in the control stage, a lightweight one-dimensional (1D) CNN classifier is employed to select the most suitable AF based on the reference signal. Notably, the classification process can be executed on a co-processor (e.g., a mobile phone), thereby reducing the computational load on the real-time controller. The real-time controller operates at the sample rate, while the co-processor functions at the frame rate, allowing both to collaborate effectively to reduce noise at the virtual microphone location.

\vspace*{-0.15cm}
\section{Transferable Selective Virtual Sensing Technique}
\vspace*{-0.1cm}
Although the CNN-based selective VS technique offers superior noise reduction performance when dealing with various primary noises compared to conventional VS techniques, it relies on noise samples from the same VS system in both the training and testing datasets. However, in practice, VS systems may differ from the one used to train the CNN model, and the pre-trained CNN model is likely to perform poorly in a new VS system. Training a separate CNN model for each system could resolve this, but it requires considerable time for data collection and model training. Moreover, this approach is impractical for systems with few or no primary noise samples.

To address this problem, a metric-learning technology ~\cite{yang2006distance,kulis2013metric} is employed to enhance the transferability of the CNN-based selective VS technique. The first step involves training a 1D CNN network to classify AFs corresponding to noises with different frequency characteristics in the training environment. 
\begin{figure}[htb]
\centering
\centerline{\includegraphics[width=8.5cm,height=7.5cm]{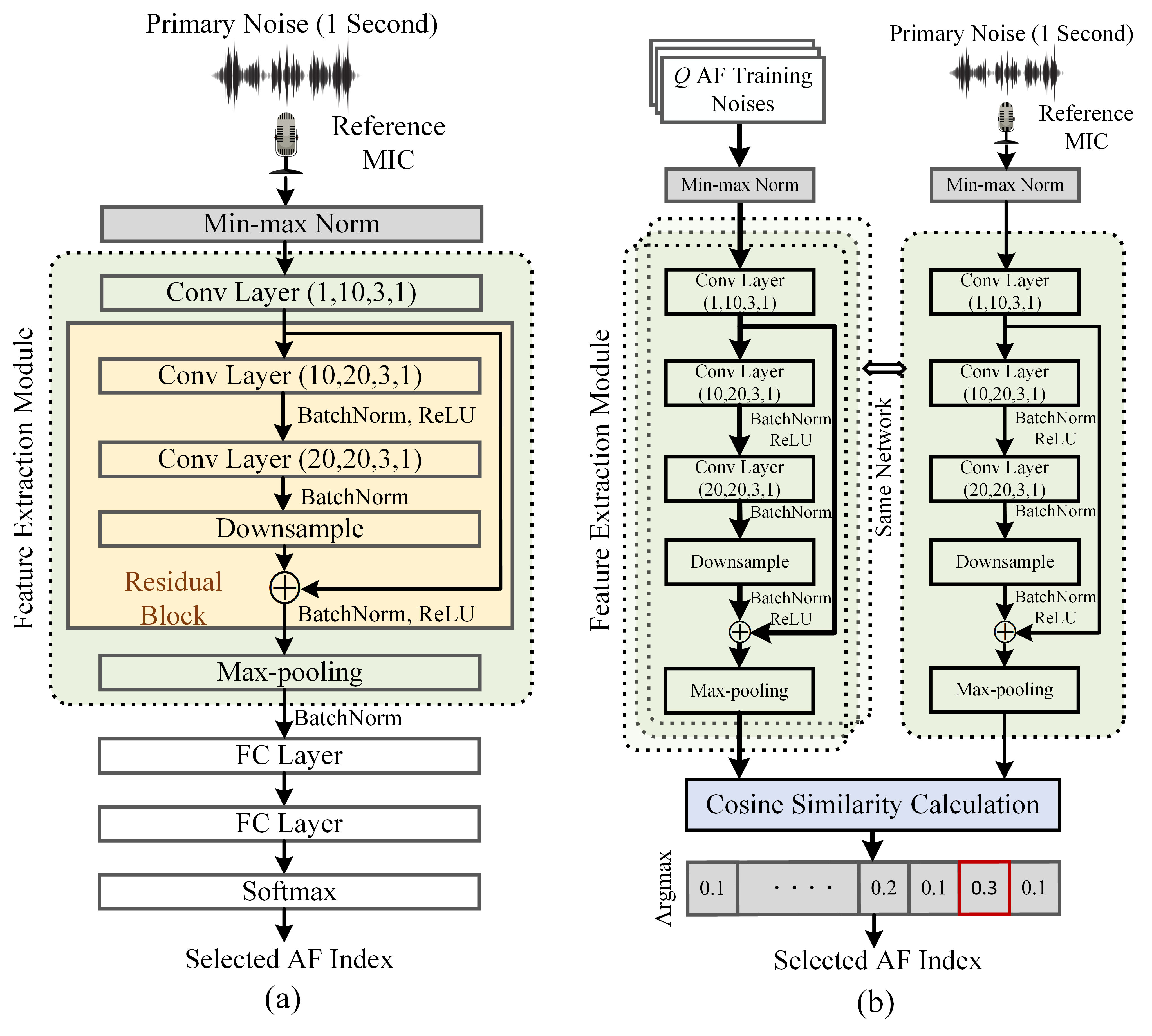}}\vspace*{-0.3cm}
\caption{Block diagram of (a) the proposed 1D CNN and (b) the classifier based on cosine similarity calculation, with each convolutional layer configured as (input channels, output channels, kernel size, and stride).}
\label{fig:2}\vspace*{-0.6cm}
\end{figure}
Fig.~\ref{fig:2} (a) illustrates the detailed architecture of the proposed 1D CNN, which features a simple yet effective structure. The input is a 1-second noise waveform, rescaled to the range of $( - 1,1)$ using min-max normalization. The preprocessed data are then passed through the feature extraction module, which consists of a convolutional layer, a residual block, and a max-pooling layer. The convolutional layers are followed by batch normalization (BN) to accelerate training and a rectified linear unit (ReLU) function to maintain positive gradients. A shortcut connection is used to add the input to the output of the residual block, simplifying the optimization process~\cite{he2016deep}. Before reaching the first fully connected (FC) layer, the max-pooling layer reduces the computational load. Finally, the output of the second FC layer is fed into the softmax layer, which calculates the probabilities for the $K$ AF classes.

Subsequently, the pre-trained 1D CNN model is employed for noise classification in the new VS system. As illustrated in Fig.~\ref{fig:2} (b), the new VS system is assumed to have $Q$ different AFs, each trained with distinct noise types. The feature extraction module of the pre-trained 1D CNN model, denoted as $\mathcal{CNN}( \cdot )$, is used to generate the embedding code for the normalized reference signal $\hat x^\prime(n)$ as
\begin{equation}
\setlength{\abovedisplayskip}{6pt}
\setlength{\belowdisplayskip}{6pt}
{{\bf{E}}_x} = \mathcal{CNN}(\hat x'(n)).
\label{Eq:6}
\end{equation}

Similarly, the embedding code of the $q$th normalized AF training noise ${{\bf{\eta }}_q}$ of the new system is derived as 
\begin{equation}
\setlength{\abovedisplayskip}{6pt}
\setlength{\belowdisplayskip}{6pt}
{{\bf{E}}_q} = \mathcal{CNN}({{\bf{\eta }}_q}) \quad q \in [1,Q].
\label{Eq:7}
\end{equation}

The cosine similarity between the embedding code of the incoming primary noise ${{\bf{E}}_x}$ and that of the $q$th AF training noise ${{\bf{E}}_q}$ can be calculated to estimate the distance distribution between the noises in the feature space as
\begin{equation}
\setlength{\abovedisplayskip}{6pt}
\setlength{\belowdisplayskip}{6pt}
{S_q} = \frac{{{\bf{E}}_x^
\mathrm{T} \cdot {{\bf{E}}_q}}}{{\max (||{{\bf{E}}_x}|{|_2} \cdot ||{{\bf{E}}_q}|{|_2},\alpha )}} \quad q \in [1,Q],
\label{Eq:8}
\end{equation}
where $\alpha$ denotes a small positive number that ensures the denominator is not zero, $\max ( \cdot )$ outputs the maximum value from the given set of inputs, and $|| \cdot |{|_2}$ refers to the L2 norm.

Finally, the selected AF is the one whose training noise shows the highest similarity to the incoming primary noise, and its index $j$ is given by
\begin{equation}
\setlength{\abovedisplayskip}{8pt}
\setlength{\belowdisplayskip}{6pt}
j = \mathop {\arg \max }\limits_{q \in [1,Q]} \{ {S_q}\}.
\label{Eq:9}
\end{equation}

The proposed transferable selective VS technique leverages the feature extraction module of the pre-trained 1D CNN model as the classifier for the new VS system, eliminating the need for retraining. Table~\ref{Table_I} compares the parameter counts between the pre-trained 1D CNN model and its feature extraction module, showing that the latter's computational complexity is about one-sixth that of the entire model. This significant reduction in computational demands enables the classifier to run efficiently on co-processors~\cite{sandler2018mobilenetv2}. Additionally, the embedding codes for the $Q$ AF training noises can be precomputed to further reduce computational load. Consequently, the proposed technique can perform AF selection on the co-processor while seamlessly interfacing with the real-time controller, providing a practical solution for VS systems.

\begin{table}[!t]
\centering \vspace*{-0.5cm}
\caption{The layers and parameters of the 1D CNN models used for Selective VS and Transferable Selective VS, respectively.}
\label{Table_I}
\vspace*{-0.3cm}
\begin{tabular}{|>{\centering\arraybackslash}m{1.5cm}|>{\centering\arraybackslash}m{1.4cm}|>{\centering\arraybackslash}m{1.0cm}|>{\centering\arraybackslash}m{1.4cm}|>{\centering\arraybackslash}m{1.0cm}|}
\hline
\multicolumn{1}{|c|}{\textbf{}} & \multicolumn{2}{c|}{\textbf{Selective VS}} & \multicolumn{2}{c|}{\textbf{Transferable Selective VS}} \\ \hline
\textbf{Layer} & \textbf{Output} & \textbf{Param} & \textbf{Output} & \textbf{Param} \\ \hline
Conv1          & [10,15998]      & 40          & [10,15998]      & 40          \\ \hline
Conv2+BN          & [20,15998]         & 640        & [20,15998]         & 640        \\ \hline
Conv3+BN      & [20,15998]         & 1240        & [20,15998]         & 1240           \\ \hline
Downsample        & [20,15998]          & 240        & [20,15998]         & 240           \\ \hline
BN      & [1,620]          & 1240         & *               & *           \\ \hline
FC1      & [1,15]          & 9315         & *               & *           \\ \hline
FC2      & [1,15]          & 240         & *               & *           \\ \hline
\textbf{Total Num} &  &    \textbf{12955}       &                 & \textbf{2160} \\ \hline
\end{tabular}\vspace*{-0.5cm}
\end{table}

\vspace*{-0.1cm}
\section{Simulation Results}
\vspace*{-0.1cm}

The effectiveness of the proposed approach is evaluated in two VS systems, the VS system \Roman{tempcounter_1} and VS system \Roman{tempcounter_2}, which differ in terms of noise types and acoustic paths. In VS system \Roman{tempcounter_1}, both the primary and secondary paths are modelled as bandpass filters with a frequency range of 20 to 7980 Hz. To simulate environmental variations, the paths in VS system \Roman{tempcounter_2} are similarly modelled but include white noise (${\rm{SNR}} = 30{\rm{dB}}$). The AFs and control filters are configured with $1024$ taps, and the systems operate at a sampling rate of $16$ kHz.


\vspace*{-0.1cm}
\subsection{Comparision of Different Networks}
\begin{figure}[!t]
\centering
\centerline{\includegraphics[width=6cm]{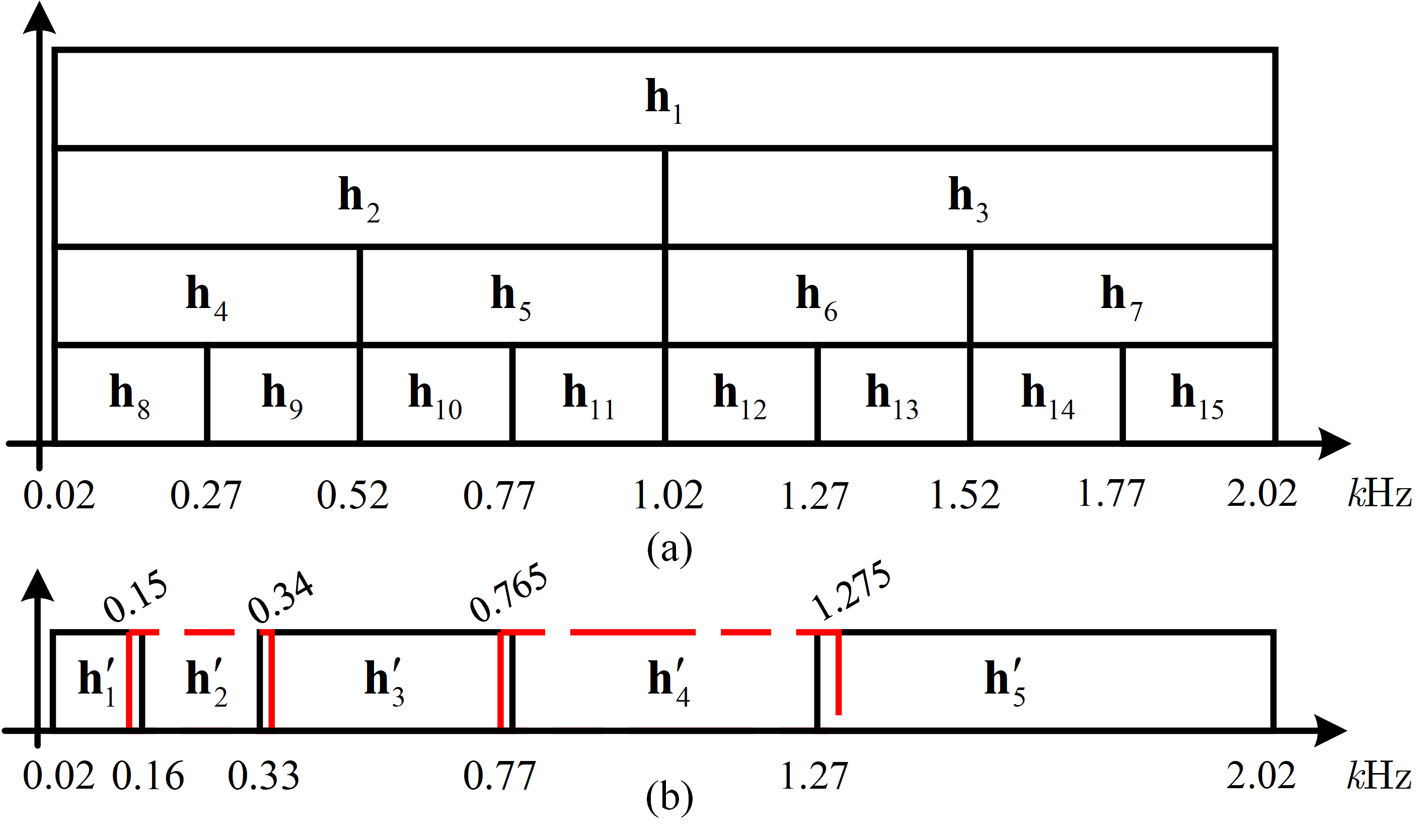}}\vspace*{-0.3cm}
\caption{The frequency spectrum of the broadband noises used to pre-train the auxiliary filters for (a) VS system \Roman{tempcounter_1} and (b) VS system \Roman{tempcounter_2}.}
\label{fig:3}\vspace*{-0.2cm}
\end{figure}

\vspace*{-0.1cm}
As illustrated in Fig.~\ref{fig:3} (a), $15$ broadband noises spanning different frequency bands are used in VS system \Roman{tempcounter_1} to pre-train the AFs. These broadband noise bands encompass the $20-2020$ Hz spectrum, focusing on the low-frequency range that ANC primarily aims to cancel. For training the 1D CNN model, $80,000$ 1-second noise waveforms with various frequency bands are synthesized for the training set, while $1,000$ 1-second waveforms are generated for both the validation and test sets. To streamline the labelling process, a frequency-band similarity ratio method is employed~\cite{shi2022selective} to automatically label the datasets by quantifying the correlation between the synthesized noise bands and those used in AF training.

Unlike in VS system \Roman{tempcounter_1}, five broadband noises with frequency ranges of $20-160$ Hz, $150-340$ Hz, $330-770$ Hz, $765-1275$ Hz, and $1270-2020$ Hz are synthesized to pre-train the AFs for VS system \Roman{tempcounter_2}, as shown in Fig.~\ref{fig:3}(b). To assess the performance of noise classification on classes not included in the prior training set, the pre-trained feature extraction module is utilized to classify $1,000$ synthesized noise samples in VS system \Roman{tempcounter_2}. Table 2 compares the noise classification accuracy of our proposed model with several 1D CNN models: M3, M5, M6-res, and M34-res~\cite{dai2017very,luo2022hybrid}. Our proposed network achieves higher noise classification accuracy in VS system \Roman{tempcounter_2} with significantly fewer parameters, indicating that a simpler network is more effective for this task and less prone to overfitting compared to more complex models.

\begin{table}[!t]
\centering \vspace*{-0.1cm}
\caption{Performance comparison of different 1D CNN models.}
\label{Table_II}
\vspace*{-0.3cm}
\begin{tabular}{|*{4}{c|}}
\hline
\textbf{Network} & \textbf{Testing Acc in \Roman{tempcounter_1}} & \textbf{Testing Acc in \Roman{tempcounter_2}} & \textbf{Param} \\ \hline
Proposed Network & 95.6\% & 92.6\% & 13K \\ \hline
M3 Network & 96.9\% & 79.9\% & 222K \\ \hline
M5 Network & 97.5\% & 63.8\% & 562K \\ \hline
M6-res Network & 98.0\% & 65.5\% & 211K\\ \hline
M34-res Network & 97.5\%  & 36.6\% & 3986K\\ \hline
\end{tabular}
\vspace*{-0.5cm}
\end{table}

\vspace*{-0.1cm}
\subsection{Varying Broadband Noise Cancellation}
\begin{figure}[!t]
    \centering
    \centerline{\includegraphics[width=7.5cm]{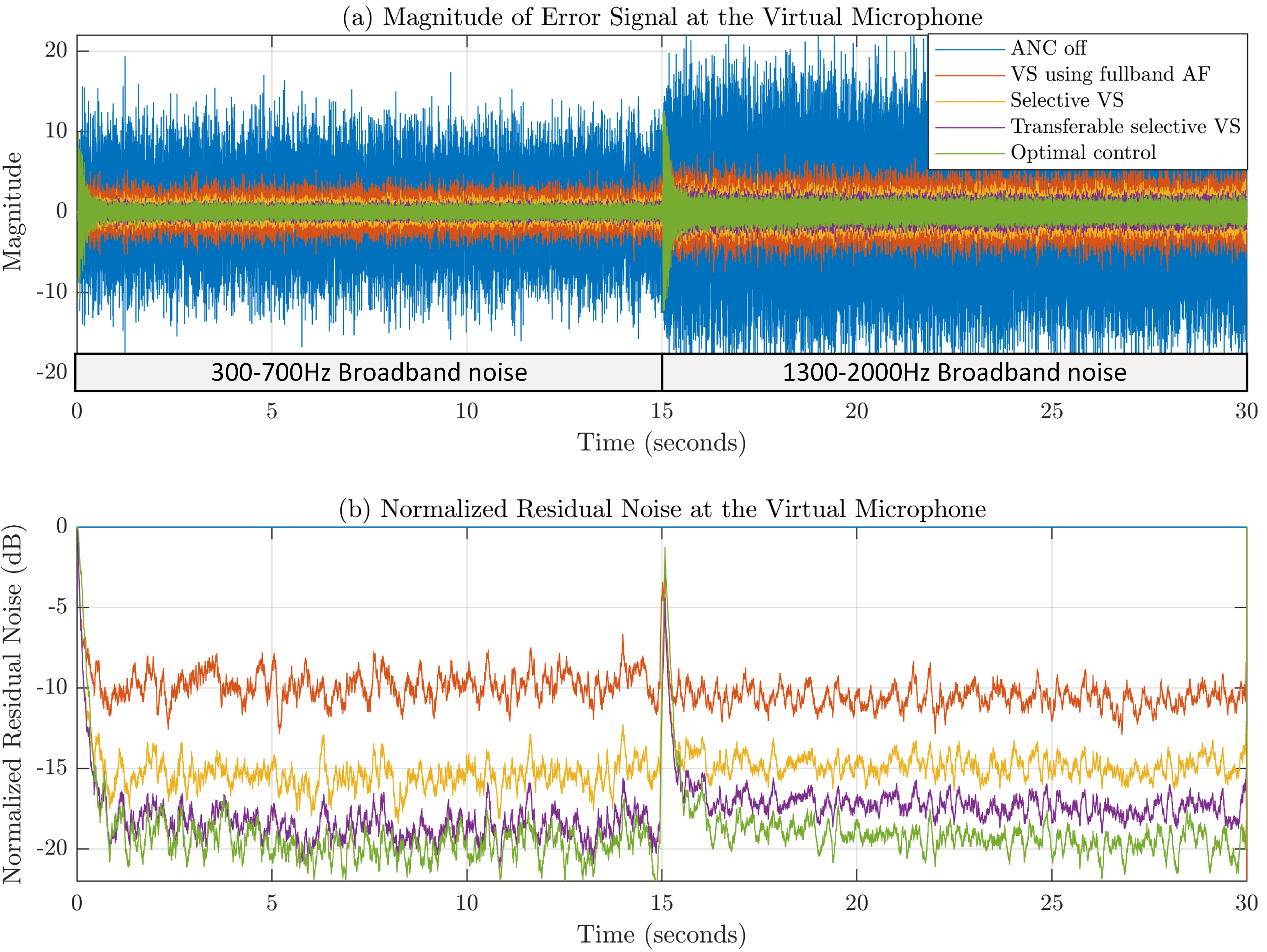}}\vspace*{-0.2cm}
    \caption{(a) Magnitude and (b) the normalized residual noise of varying broadband noise attenuated by different algorithms at the virtual microphone.}\vspace*{-0.2cm}
    \label{fig:4}
\end{figure}

\vspace*{-0.1cm}
In this simulation, the performance of the transferable selective VS technique in system \Roman{tempcounter_2} is evaluated under varying broadband noise conditions. The varying primary noise is generated by combining two $15$-second broadband noises, one spanning $300-700$ Hz and the other $1300-2000$ Hz. During the simulation, the proposed method adapts the AF every second, with the selected AF index sent to the real-time controller for noise reduction. Fig.~\ref{fig:4} illustrates the time history of the error signal and the normalized residual noise~\cite{li2023distributed} at the virtual microphone. The noise reduction performance of the transferable selective VS technique is compared to that of the VS technique using a full-band AF ($20-2020$ Hz), the selective VS technique, and the optimal control where the FxLMS algorithm directly cancels the error signal captured by the virtual error microphone. The results demonstrate that the proposed transferable selective VS technique achieves superior noise reduction performance, closely matching the optimal control at the virtual position and outperforming both the full-band AF and the selective VS technique. This simulation confirms that the proposed algorithm can effectively operate under varying broadband noise conditions.
\begin{figure}[!t]
    \centering
    \centerline{\includegraphics[width=7.5cm]{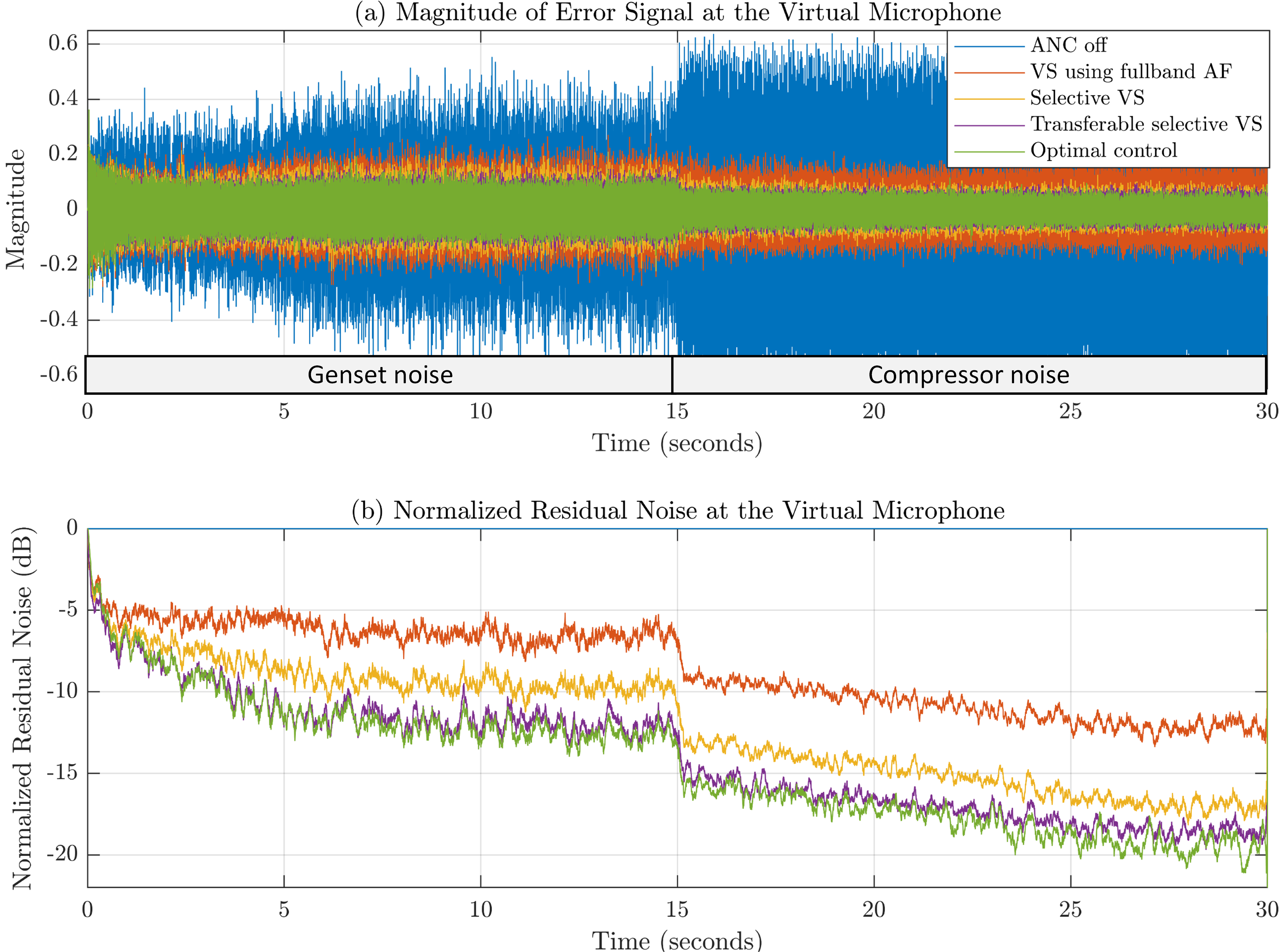}}\vspace*{-0.2cm}
    \caption{(a) Magnitude and (b) the normalized residual noise of varying real-world noise attenuated by different algorithms at the virtual microphone.}\vspace*{-0.55cm}
    \label{fig:5}
\end{figure}

\vspace*{-0.1cm}
\subsection{Real-world Noise Cancellation}
\vspace*{-0.1cm}
In this simulation, the proposed transferable selective VS technique is applied to attenuate real-world noises. The varying noise is generated by concatenating 15-second recordings of genset noise and compressor noise, neither of which are included in the training set. The time history of the virtual error signal in Fig.~\ref{fig:5} shows that the AFs selected by the pre-trained feature extraction module effectively reduce disturbances at the desired virtual location. Furthermore, the proposed transferable selective VS technique outperforms both the VS with the full-band AF and the selective VS technique in noise reduction, achieving performance nearly on par with optimal control. The results validate the effectiveness of the proposed method in handling varying real-world noises that cannot be effectively managed by conventional selective VS techniques lacking transferability.

\vspace*{-0.3cm}
\section{Conclusion}
\vspace*{-0.2cm}
In this paper, we proposed the transferable selective VS technique to enhance the transferability of CNN-based selective VS methods. Using metric-learning technology, the proposed approach leverages a pre-trained feature extraction module to achieve noise classification in new VS systems. This technique allows the CNN classifier to be trained on one VS system and then applied to another without requiring retraining, with its low parameter count ensuring efficient operation on co-processors. Despite requiring significantly fewer parameters than more complex networks, the proposed 1D CNN network achieves higher accuracy in distinguishing various noise classes. The simulation results demonstrate the effectiveness of our method in reducing both sudden-varying broadband noises and real-world noises.

\newpage
\bibliographystyle{IEEEbib}
\bibliography{refs}

\end{document}